\documentclass[journal=jctcce,manuscript=article]{achemso}
\usepackage{amsmath}
\usepackage{color}
\usepackage{graphicx}
\usepackage{placeins}
\usepackage{subfigure}

\title{Low-Scaling Many-Body Green's Function Calculations for Molecular Systems via Interacting-Bath Dynamical Embedding Theory}
\author{Christian Venturella}
\author{Jiachen Li}
\author{Tianyu Zhu}
\email{tianyu.zhu@yale.edu}
\affiliation{Department of Chemistry, Yale University, New Haven, CT, USA 06520}

\begin{document}

\begin{abstract}
We present a molecular extension of our recently proposed Green's function embedding method, interacting-bath dynamical embedding theory (ibDET), for computing charged excitation energies at the $GW$ and EOM-CCSD levels. Starting from atom-centered impurities, we construct bath representations that capture the frequency-dependent entanglement between the impurity and its environment and can be systematically improved via the construction of cluster-specific natural orbitals. Utilizing a $GW$ or coupled-cluster Green's function solver, the self-energy of the full system is assembled from all embedding problems to obtain the interacting Green's function. We show that ibDET provides accurate spectral properties with much reduced cost for a broad range of systems, including conjugated molecules and nanoclusters. Compared with full-system results, the errors in the predicted ionization potentials and electron affinities are around 0.1 eV or smaller, while each embedding problem includes only a small fraction of the total orbital space. This work provides an efficient and scalable framework for computing spectral properties of molecular systems.
\end{abstract}

\maketitle

\section{Introduction}

First-principles modeling of molecules and materials in their excited states is central to chemistry and materials science. Computational methods can provide predictions with physical insights, thus aiding in the design of new platforms for energy harvesting, information storage, sensing, and catalysis~\cite{onidaElectronicExcitationsDensityfunctional2002,norskovDensityFunctionalTheory2011,zhuChargeTransferMolecular2018}. 
Density functional theory (DFT)~\cite{kohnSelfConsistentEquationsIncluding1965,parrDensityFunctionalTheoryAtoms1989} has been the most popular approach to simulate the electronic structure of molecules, liquids, and solids, due to the excellent balance between accuracy and computational cost.
However,
it is well-known that the Kohn-Sham DFT framework with semi-local functionals often predicts spectral properties with large errors because of the delocalization error\cite{cohenInsightsCurrentLimitations2008}.
In addition,
DFT suffers from the undesired dependence on the choice of the exchange-correlation (XC) functional.

To simulate response properties more accurately than DFT,
Green's function methods derived from Hedin's equations\cite{hedinNewMethodCalculating1965} have become increasingly prominent in recent years.
In the Green's function formalism,
the poles of the one-particle Green's function along the frequency domain have a clear physical meaning as the charged excitation quasiparticle energies of electron removal and addition processes, and directly measurable in photoemission spectroscopy \cite{reiningGWApproximationContent2018,golzeAccurateAbsoluteRelative2020}. Among Green's function methods, many-body perturbation theories, such as the $GW$ approach, constitute the lowest-order approximations to Hedin's equations.
In the $GW$ approach,
the bare Coulomb interaction is replaced by a dynamically screened interaction to describe electron and hole propagation in a weakly interacting system, which leads to improved performance in predicting charged excitation energies over DFT\cite{golzeCoreLevelBindingEnergies2018,golzeGWCompendiumPractical2019,liuDielectricEmbeddingGW2020,wilhelmLowScalingGWBenchmark2021,zhuAllElectronGaussianBasedG0W02021,liBenchmarkGWMethods2022,liRenormalizedSinglesCorrelation2022,leiGaussianbasedQuasiparticleSelfconsistent2022,moninoConnectionsPerformancesGreens2023,liuManyBodyEffectsHeterogeneous2025,zhouAllElectronBSEGWMethod2025}.
Furthermore,
from $GW$ quasiparticle energies,
accurate optical spectra can be obtained using the Bethe-Salpeter equation (BSE) formalism\cite{blaseBetheSalpeterEquation2020,moninoSpinConservedSpinFlipOptical2021,liCombiningLocalizedOrbital2022,liCombiningRenormalizedSingles2022,bhattacharyaBSEGWPredictionCharge2024,hillenbrandEnergyspecificBetheSalpeter2025}.

To achieve a quantitative description and reduced starting-point dependence,
highly accurate many-body theories beyond low-order perturbation theories are needed to construct the Green's function,
including the coupled-cluster (CC) theory\cite{nooijenCoupledClusterApproach1992,nooijenCoupledClusterGreens1993,pengGreensFunctionCoupledCluster2018,zhuCoupledclusterImpuritySolvers2019,sheeCoupledClusterImpurity2019,laughonPeriodicCoupledClusterGreens2022}, 
algebraic diagrammatic construction (ADC) theory\cite{banerjeeThirdorderAlgebraicDiagrammatic2019},
density matrix renormalization group (DMRG)\cite{roncaTimeStepTargetingTimeDependent2017},
and quantum Monte Carlo\cite{gullContinuoustimeMonteCarlo2011}.
These approaches provide a systematically improvable path to go beyond the $GW$ approximation,
which only includes ring diagrams through the random phase approximation (RPA)\cite{martinInteractingElectrons2016}.
For example,
the accuracy of the coupled-cluster Green's function (CCGF) that is equivalent to the equation-of-motion coupled-cluster (EOM-CC) formulation can be systematically improved by including high-order diagrams.
Despite the excellent accuracy,
their applications to large-scale systems remain computationally prohibited. To address this challenge, many methods have emerged to reduce simulation costs. 
For $GW$ methods, low-scaling formulations have been developed to reduce bottlenecks associated with the RPA polarizability and dielectric function using techniques such as resolution of the identity (RI), integral screening, and stochastic sampling \cite{wilhelmLowScalingGWBenchmark2021,amblardManybodyGWCalculations2023,allenGWHybridFunctionals2024,yehLowScalingAlgorithmsGW2024}. Additionally, low-scaling methods have been developed for CC theory for modeling both the ground-states and charged excited-states of molecules \cite{duttaNearlinearScalingEquation2018,duttaDomainbasedLocalPair2019}.
More recently,
data-driven machine learning has been used to accelerate many-body Green's function calculations\cite{venturellaMachineLearningManyBody2024,dongEquivariantNeuralNetwork2024,venturellaUnifiedDeepLearning2025}.

Quantum embedding offers another promising pathway to low-scaling predictive calculations.
In quantum embedding methods,
a local subsystem of interest is treated with an accurate but computationally demanding high-level theory,
while the remaining system is described by a 
low-level theory to provide effective potential or interaction\cite{sunQuantumEmbeddingTheories2016,martinInteractingElectrons2016,mejuto-zaeraQuantumEmbeddingMolecules2024,mejutozaeraGhostEmbedding2026}.
Much progress has been made in 
density matrix embedding theory (DMET)\cite{kniziaDensityMatrixEmbedding2012,fertittaEnergyweightedDensityMatrix2019,cuiEfficientImplementationInitio2020},
dynamical mean-field theory (DMFT)\cite{georgesHubbardModelInfinite1992,georgesDynamicalMeanfieldTheory1996,zhuEfficientFormulationInitio2020,zhuInitioFullCell2021,zgidDynamicalMeanfieldTheory2011}, and
self-energy embedding theory (SEET)\cite{lanCommunicationInitioSelfenergy2015,rusakovSelfEnergyEmbeddingTheory2019}, which achieved success for quantitative predictions of electronic properties for a broad range of systems including molecules, correlated materials, and point defects\cite{sunExtendedDynamicalMeanfield2002,biermannFirstPrinciplesApproachElectronic2003,sunFinitetemperatureDensityMatrix2020,liRestoringTranslationalSymmetry2024,zhuExactElectronicQuantum2025}.
Particularly, DMFT is a Green's function embedding method designed to treat strongly correlated systems such as transition metal materials and lanthanides with localized atomic-like states. However, extending DMFT to accurately capture nonlocal and long-range electron correlation remains challenging for predicting spectral properties of multi-fragment, molecular systems\cite{reuterRealSpaceElectron2021,yuDescribingChemicalReactivity2022}.

Recently, we developed a new Green's function embedding approach,
interacting-bath dynamical embedding theory (ibDET)\cite{liInteractingBathDynamicalEmbedding2024},
to simulate spectral properties of solid-state materials.
In ibDET,
the impurity problem is constructed in a local orbital basis,
then the embedding space is expanded by selecting bath orbitals that entangle most strongly with impurity orbitals from the environment.
Compared to traditional Green's function embedding approaches (e.g., DMFT and SEET) that construct non-interacting bath orbitals through the hybridization function\cite{sunQuantumEmbeddingTheories2016,martinInteractingElectrons2016},
the interacting bath orbitals in ibDET capture two-particle interactions in a systematically improvable manner,
which avoids uncontrolled errors associated with small impurity subspace and empirical truncations. This is achieved by gradually expanding the embedding space with cluster-specific natural orbitals whose occupancies exceed a tunable, predefined threshold, thereby allowing the embedding space to capture long-range interactions. 
With the interacting bath orbitals,
the full Hamiltonian can be directly projected to the embedding space.
Because the bath orbitals properly describe both short-range and long-range interactions,
ibDET successfully captures local and nonlocal electron correlations on the equal footing. As shown in our previous work,
ibDET provides good agreement of photoemission spectra compared with experiments for a range of moderately correlated insulators, semiconductors, and metals\cite{liInteractingBathDynamicalEmbedding2024}.
In this work, we develop an extension of ibDET for molecular systems, where the goal is to accelerate $GW$ and CC Green's function calculations. 
We demonstrate the method's good balance between efficiency and accuracy for selected molecular and nanomaterial systems. 
We show that the full-space results can usually be obtained with less than 300 embedding orbitals,
corresponding to a small fraction of total orbital space.
For both $GW$ and EOM-CCSD,
the errors of calculating IPs and EAs are around $0.1$ eV or smaller compared to full-space results.

\section{Methods}
\subsection{Green's Function Formalism}
The one-particle Green's function describes propagation of particle (p) and hole (h) states in the many-electron system. In the frequency domain, the Green's function can be split into addition part $G^+(\omega)$ and removal part $G^-(\omega)$ as:
\begin{subequations}
\begin{align}
 G^+_{pq}(\omega) &=  \langle \Psi_0 | \hat{a}_p  
   [\omega - (\hat{H}-E + i\eta) ]^{-1}  \hat{a}^\dagger_q | \Psi_0 \rangle , \\
  G^-_{pq}(\omega) &=  \langle \Psi_0 | \hat{a}^\dagger_q  
    [\omega + (\hat{H}-E - i\eta) ]^{-1}  \hat{a}_p | \Psi_0  \rangle,
\end{align}
\label{eq:exactgf}%
\end{subequations}
where $|\Psi_0\rangle$ is the ground-state wave function, $\hat{H}$ is the Hamiltonian, $E$ is the ground-state energy, 
$\hat{a}_p$ and $\hat{a}_q^\dagger$ are annihilation and creation operators,
and $\eta$ is the broadening parameter. 
In this paper, we use $i$, $j$, $k$, $l$ for occupied orbitals, 
$a$, $b$, $c$, $d$ for virtual orbitals, 
$p$, $q$, $r$, $s$ for general orbitals.

From the Green's function,
the spectral function $A$ and electronic density of states (DOS) can be obtained by
\begin{subequations}
    \begin{align}
        A(\omega) \equiv & -\frac{1}{\pi} \text{Im} \left[G(\omega)\right], \\
        \text{DOS} (\omega) \equiv &  ~\text{Tr} A(\omega).
    \end{align}
\label{eq:dos}%
\end{subequations}
which can be directly measured in photoemission and inverse photoemission spectroscopies.
Starting from a mean-field Green's function $G_0$,
the many-body Green's function $G$ can be calculated via the Dyson equation
\begin{equation}\label{eq:dyson}
    \Sigma(\omega) = G_0^{-1}(\omega) - G^{-1}(\omega),
\end{equation}
where $\Sigma(\omega)$ is the self-energy describing electron correlation effects.

\subsection{Interacting-Bath Dynamical Embedding Theory}
\FloatBarrier
We then describe the Green's function embedding framework in ibDET.
For a given molecule,
ibDET starts with the mean-field solution using Gaussian atomic orbitals.
Then the impurity problem is defined in the intrinsic atomic orbital plus projected atomic orbital (IAO+PAO)\cite{kniziaDensityMatrixEmbedding2013,cuiEfficientImplementationInitio2020} space,
which consists of orthogonal, atom-centered orbitals local in the real space. The IAOs capture the valence space, while the PAOs capture the remaining higher virtual space. Together this IAO+PAO basis covers the same space as the original AO basis.
In this work,
we choose local orbitals of each non-hydrogen atom plus its bonded hydrogen atoms as an impurity fragment,
then gradually expand the bath space with a multitier scheme to capture the short- and long-range electron correlation. Since we construct multiple atom-centered impurity problems, each having its own nonlocal embedding space, the resulting embedding problems naturally overlap.

First,
employing the idea of DMET\cite{kniziaDensityMatrixEmbedding2013,cuiEfficientImplementationInitio2020},
bath orbitals $B_\mathrm{DM}$ that exactly reproduce the mean-field one-particle reduced density matrix (1-RDM) of the impurity are included,
which are obtained from the singular value decomposition (SVD) of the mean-field off-diagonal 1-RDM $\gamma$ between the impurity and the environment 
\begin{equation}\label{eq:bath_dm}
    \gamma^{\text{imp,env}} = B_{\text{DM}} \Lambda V^{\dagger}
\end{equation}

\begin{figure}[hbt!]
    \centering
    \includegraphics[width=0.7\linewidth]{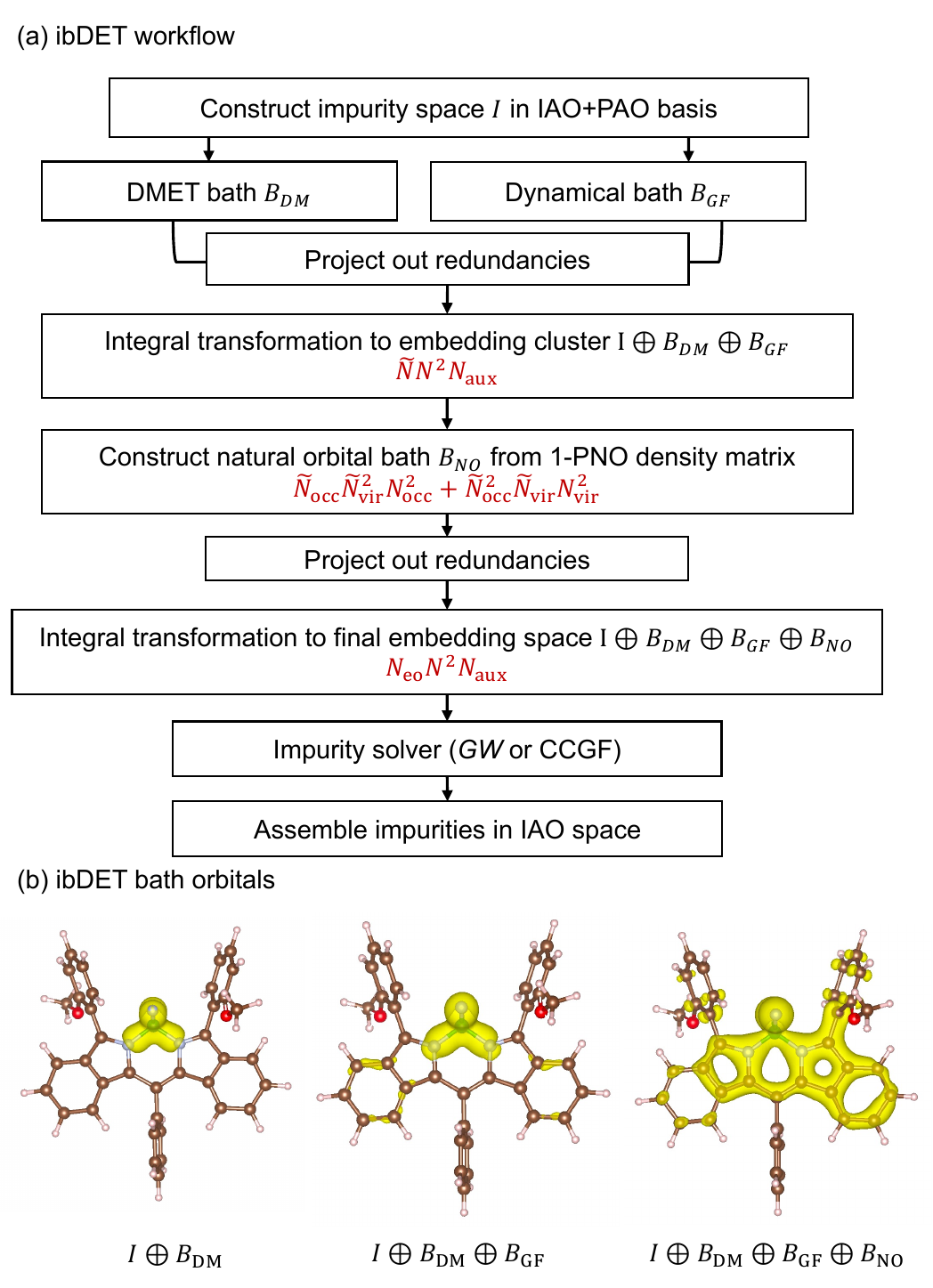}
    \caption{ (a) Illustration of the ibDET approach for molecules. 
    Scaling bottlenecks associated with ibDET integral transformations are given in red. 
    (b) An example of systematically increasing occupied embedding space in BODIPY. Showing the embedding space electron density.}
    \label{fig:1}
\end{figure}

Secondly, 
bath orbitals $B_{\text{GF}}$ that reproduce the mean-field Green's function of the impurity are constructed~\cite{nusspickelEfficientCompressionEnvironment2020}.
Inspired by the relation between the 1-RDM and the one-body Green's function at the static limit,
we discretize the mean-field one-body Green's function on a uniform set of real-axis frequency points $\{\omega_n\}$ for the dynamical entanglement between the impurity and the environment.
The occupied and virtual parts of the one-body mean-field Green's function are defined as
\begin{subequations}
    \begin{align}
        (G_0^\text{occ,MO})_{ii} (\omega_n) = & 
        \frac{1}{\omega_n - \epsilon_i + i \eta} \\ 
        (G_0^\text{vir,MO})_{aa} (\omega_n) = & 
        \frac{1}{\omega_n - \epsilon_a + i \eta}
    \end{align}
\end{subequations}
where $\epsilon$ is the orbital energy.
After rotating occupied and virtual parts of the mean-field Green's function $G_0^{\text{MO}}$
into the local orbital space by the transformation matrix $C^{\text{MO,LO}}$,
the occupied and virtual bath orbitals $B_{\text{GF}}$ can be obtained from the SVD of the imaginary part of the off-diagonal $G_0^\text{occ,LO}$ and $G_0^\text{vir,LO}$
\begin{subequations}
    \begin{align}
        \text{Im} [G_0^\text{occ,LO}]^\text{imp,env} (\omega_n) 
        = & \left [ B_\text{GF}^\text{occ} \Lambda V^{\dagger} \right ] (\omega_n)\\
        \text{Im} [G_0^\text{vir,LO}]^\text{imp,env} (\omega_n) 
        = & \left [ B_\text{GF}^\text{vir} \Lambda V^{\dagger} \right ] (\omega_n)
    \end{align}
\end{subequations}
where the bath orbitals $B_{\text{GF}}$ are assembled as
\begin{equation}
    B_{\text{GF}} = \left [ 
    B_\text{GF}^\text{occ} (\omega_1), B_\text{GF}^\text{vir} (\omega_1), 
    B_\text{GF}^\text{occ} (\omega_2), B_\text{GF}^\text{vir} (\omega_2), 
    \text{...}
    \right ]
\end{equation}
$B_{\text{GF}}$ can be considered as the dynamical extension to the static $B_{\text{DM}}$\cite{nusspickelEfficientCompressionEnvironment2020}.
Note that bath orbitals $B_{\text{GF}}$ from discretizing the mean-field Green's function are not orthogonal.
To remove the redundancy,
a projection step is applied to remove the embedding orbitals that overlap minimally with the full-system Hilbert space and orthogonalize the embedding space~\cite{liInteractingBathDynamicalEmbedding2024}.
As shown in Fig.~\ref{fig:1},
the embedding space $I \bigoplus B_\mathrm{DM} \bigoplus B_\mathrm{GF}$ ($I$ is impurity space) spans a local real space around the impurity atom,
which covers the short-range and medium-range electron correlations. 

To further capture the long-range electron correlation,
the cluster-specific natural orbitals\cite{nusspickelSystematicImprovabilityQuantum2022} are utilized to expand the existing embedding space.
In this work,
we introduce an improved scheme over our previous work~\cite{liInteractingBathDynamicalEmbedding2024} for constructing cluster-specific MP2 density matrix to select bath orbitals with strongest couplings to the embedding space,
which shares the idea of pair natural orbitals (PNOs)\cite{meyerPNOCIStudies1973,ahlrichsPNOCIPair1975} in local correlation approaches
\begin{subequations}
\begin{align}
    \gamma_{ij} = & 2 \delta_{ij} - 2 \sum_{\tilde{k}\tilde{a}\tilde{b}} t^{\tilde{a}\tilde{b}}_{i\tilde{k}} \left [ 2 t^{\tilde{a}\tilde{b}}_{j\tilde{k}} - t^{\tilde{b}\tilde{a}}_{j\tilde{k}} \right ] \\
    \gamma_{ab} = & 2 \sum_{\tilde{i}\tilde{j}\tilde{\tilde{c}}} t^{a\tilde{c}}_{\tilde{i}\tilde{j}} \left [  2 t^{b\tilde{c}}_{\tilde{i}\tilde{j}} - t^{\tilde{c}b}_{\tilde{i}\tilde{j}} \right ]
\end{align}
\label{eq:1pno}%
\end{subequations}
In Eq.\ref{eq:1pno},
the MP2 amplitudes are calculated as
\begin{subequations}\label{eq:mp2_amplitude}
    \begin{align}
        t_{i\tilde{j}}^{\tilde{a}\tilde{b}} = & -\frac{(i\tilde{a}|\tilde{j}\tilde{b})}{\epsilon_{\tilde{a}} + \epsilon_{\tilde{b}} - \epsilon_i - \epsilon_{\tilde{j}}},\\
        t_{\tilde{i}\tilde{j}}^{a\tilde{b}} = & -\frac{(\tilde{i}a|\tilde{j}\tilde{b})}{\epsilon_a + \epsilon_{\tilde{b}} - \epsilon_{\tilde{i}} - \epsilon_{\tilde{j}}}
    \end{align}
\end{subequations}
The tilde over an index indicates that the corresponding orbital belongs to the embedding cluster $I \bigoplus B_\mathrm{DM} \bigoplus B_\mathrm{GF}$, whereas an index without a tilde denotes an orbital in the environment.  Although the use of cluster-specific natural orbitals is conceptually similar to the embedding method in Ref.~\citenum{nusspickelSystematicImprovabilityQuantum2022}, the construction and role of these orbitals in ibDET differ in two important aspects. First, the embedding cluster in Ref.~\citenum{nusspickelSystematicImprovabilityQuantum2022} is constructed to reproduce the impurity density matrix, whereas the embedding cluster in ibDET also incorporates the dynamical bath orbitals $B_{\mathrm{GF}}$ to reproduce the mean-field Green's function on the impurity. Second, the present work employs a different formulation of the MP2 amplitudes for constructing the cluster-specific natural orbitals, which significantly reduces the computational cost.
In Ref.\citenum{nusspickelSystematicImprovabilityQuantum2022} and our previous work\cite{liInteractingBathDynamicalEmbedding2024},
two indexes of MP2 amplitudes are in the environment. In this work, we propose a new scheme where 
only one index of MP2 amplitudes is in the environment, as shown in Eq.~\ref{eq:mp2_amplitude}. Thus, we denote this scheme as ``1-PNO''.
This 1-PNO scheme significantly reduces the computational cost to evaluate the MP2 density matrix, while maintaining similarly consistent convergence behavior with respect to the embedding size. The number of environment indices (1-PNO vs. 2-PNO) defines a systematic hierarchy for constructing the correlated bath. The 1-PNO scheme is the lowest-order member of this hierarchy that assigns MP2 occupation or hole weight to individual environment orbitals. Our one-environment-index construction provides the most direct criterion for selecting environment orbitals that are correlated with the embedding cluster.
Including two or more environment indices would capture higher-order environment-pair correlations, but at a cost that scales quadratically or higher with the size of the residual environment. Instead, we control the bath incompleteness by systematically increasing the embedding size.
The dominant step that constructs MP2 amplitudes in the 1-PNO scheme scales as $\mathcal{O} (N_\text{aux} N^2 \tilde{N}_\text{occ})$ + $\mathcal{O} (N_\text{aux} N^2 \tilde{N}_\text{vir})$,
where $N_\text{aux}$ and $N$ are the numbers of auxiliary and atomic basis functions in the full system, and
$\tilde{N}_\text{occ}$ and $\tilde{N}_\text{vir}$ are the numbers of occupied and virtual orbitals in the embedding cluster. The step that computes cluster-specific MP2 density matrices scales as $\mathcal{O}(\tilde{N}_\text{occ}\tilde{N}_\text{vir}^2 N_\text{occ}^2) + \mathcal{O}(\tilde{N}_\text{occ}^2\tilde{N}_\text{vir} N_\text{vir}^2)$, where $N_\text{occ}$ and $N_\text{vir}$ are the numbers of occupied and virtual orbitals in the full system. Then, these density matrices are diagonalized to obtain the occupied and virtual cluster-specific natural orbitals $B_\text{NO}$, and only those with the largest fractional occupancies above a chosen threshold are retained. These $B_\text{NO}$ incorporate nonlocal correlations into the bath space, as shown via the embedding space electron density in Fig. \ref{fig:1}b. We define $N_\mathrm{eo}$ as the number of orbitals in the bath space after the final projection step.

Fig.~\ref{fig:1} provides more algorithmic details of our molecular ibDET method, with computational scalings for key intermediate steps. 
Bath orbitals $B_{\text{DM}}$ and $B_{\text{GF}}$ are determined based on the mean-field calculations.
Thus, the size of the embedding space is mainly controlled via thresholds for selecting occupied and virtual cluster-specific natural orbitals $B_{\text{NO}}$. 
After deriving the embedding problems,
a Green's function solver is used to obtain the Green's function and the self-energy in the embedding space at the level of many-body perturbation theory ($GW$) or CC theory (EOM-CCSD). 
We emphasize that $B_\mathrm{NO}$ is not constructed from a screened interaction, such as the screened interaction derived from constrained RPA (cRPA). 
In the present formulation, the MP2-level density matrix is used only as a correlated orbital-selection metric to identify environment orbitals most strongly coupled to the embedding cluster. 
After the embedding space $I\bigoplus B_\mathrm{DM}\bigoplus B_\mathrm{GF}\bigoplus B_\mathrm{NO}$ is defined, 
the full Hamiltonian with bare two-electron interaction is projected into this space and solved using the chosen impurity Green's function solver. 
Therefore,
the limit of ibDET is achieved through a systematically improvable orbital bath.

As shown in Refs.~\citenum{renResolutionofidentityApproachHartree2012,zhuAllElectronGaussianBasedG0W02021},
the scaling of evaluating the $GW$ self-energy is $N_\omega N_{\text{occ}} N_{\text{vir}} N^2_{\text{aux}}$
where $N_\omega$ is the number of frequency points.
For the $GW$ embedding problem,
the three-center density-fitting integral is obtained through the Cholesky decomposition of the electron repulsion integrals (ERI) in the embedding space as $(pq|rs) = \sum_P v^{pq}_P v^{rs}_P$.
Compared to $GW$, CCGF admits a much higher overall scaling. Solving the CCSD ground-state carries a scaling of $\mathcal{O}(N_\text{occ}^2 N_\text{vir}^4)$. while
solving the IP and EA parts of the full CCGF matrix scales as $\mathcal{O}(N_\omega N_\text{occ} N_\text{vir}^3 N^2)$.
After solving the CCGF,
the CC self-energy is obtained from the Dyson equation in Eq.~\ref{eq:dyson}. \par
To finally assemble the self-energy of the whole system from multiple impurity self-energies, the self-energy of the $I$-th impurity in the embedding space $\Sigma^\text{emb,I}$ is rotated to the full space $\Sigma^\text{full,I}$.
Then the diagonal block of the full-system self-energy is taken as $\Sigma^\text{full,I}$ of the corresponding impurity, 
and the off-diagonal block between the $I$-th and $J$-th impurity is assembled via democratic partitioning
\begin{equation}
\Sigma^\text{full}_{pq} = \frac{1}{2}\!\left(\Sigma^\text{full,I}_{pq} + \Sigma^{\text{full,J}}_{pq}\right)
\end{equation}

\section{Computational Details}
We benchmarked molecular ibDET for the ionization potentials (IPs) and electron affinities (EAs) of four molecular systems,
including a silicon nanocluster, hydrogenated phosphorene nanosheets, quaterrylene, and BODIPY. 
For the silicon nanocluster and phosphorene nanosheets,
the cc-pVTZ basis set\cite{dunningGaussianBasisSets1989} was used.
For quaterrylene and BODIPY,
the cc-pVDZ basis set\cite{dunningGaussianBasisSets1989} was used.
Geometries can be found in the SI.
All ground-state Hartree-Fock (HF) and IP/EA-EOM-CCSD calculations were performed using the PySCF quantum chemistry software package\cite{sunPySCFPythonbasedSimulations2018,sunRecentDevelopmentsPySCF2020,sun2026python}.
Full-space $G_0W_0$ calculations based on the HF reference were performed with the fcDMFT package\cite{zhuEfficientFormulationInitio2020,zhuAllElectronGaussianBasedG0W02021,zhuInitioFullCell2021}.

In ibDET calculations, the MINAO basis set was used as the reference basis for constructing IAOs.
Local orbitals of each non-hydrogen atom and its bonded hydrogen atoms were selected as an impurity.
In the frequency-dependent bath orbital $B_{\text{GF}}$ construction,
uniform real-frequency grids between 0.3 a.u. below the HOMO energy and 0.3 a.u. above the LUMO energy were used,
where the broadening parameter was 0.03 a.u.
In the natural bath orbital construction,
occupied natural orbitals with occupation numbers larger than a user-defined threshold were selected,
then the number of virtual natural orbitals was determined as $N_{\text{vir,NO}}=4 N_{\text{occ,NO}}$.
As shown in the SI,
different tested $N_{\text{vir,NO}}: N_{\text{occ,NO}}$ ratios give similar quasiparticle results.
To improve the ibDET convergence with respect to the embedding size,
ten low-lying canonical virtual orbitals were always added to the embedding space.
After constructing embedding problems,
the $GW$ or CCGF solver are applied to compute the self-energy,
and the resulting ibDET levels are denoted as HF+$GW$ or HF+CC. After obtaining the full-space Green's function,
the quasiparticle energies are fitted from the DOS.

\section{Results}
\subsection{Nanomaterials with $GW$ Impurity Solver}
\begin{figure}[hbt!]
    \centering
    \includegraphics[width=0.8\linewidth]{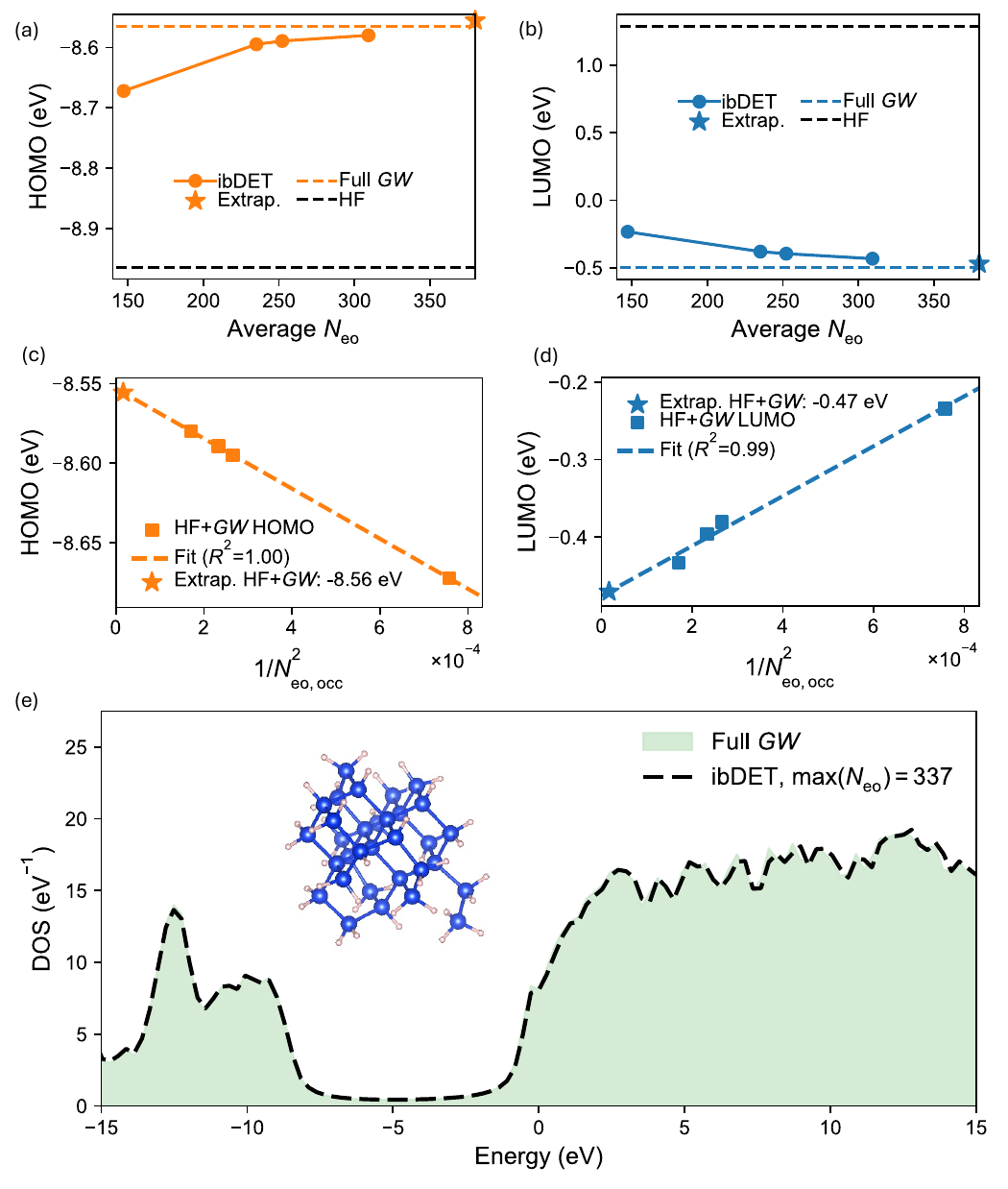}
    \caption{Silicon nanocluster HF+$GW$ ibDET benchmark. (a,b) HOMO, LUMO quasiparticle energies vs. embedding size, with Hartree-Fock and full $G_0W_0$@HF for reference. (c,d) Extrapolation to the full-space limit with respect to embedding size. Extrapolated values shown on (a) and (b) as a star. The full-space $GW$ HOMO and LUMO are $-8.57$ eV and $-0.50$ eV respectively. (e) ibDET density of states compared against full-space $G_0W_0$@HF.}
    \label{fig:2}
\end{figure}

\begin{figure}[hbt!]
    \centering
    \includegraphics[width=0.7\linewidth]{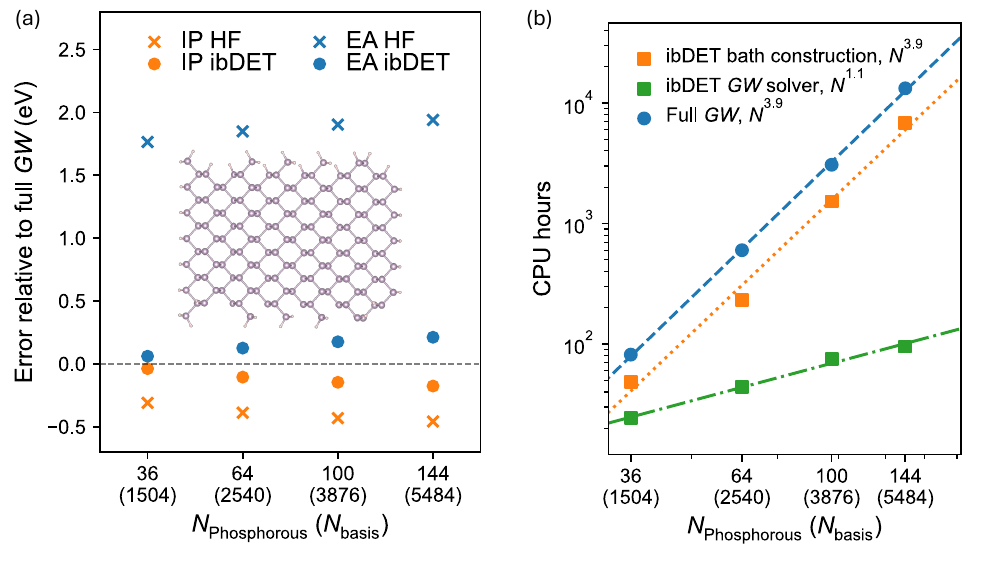}
    \caption{Phosphorene nanosheet ibDET HF+$GW$ benchmark. (a) Errors for HOMO and LUMO quasiparticle energies relative to full-space results vs. nanosheet size. Largest nanosheet ($6\times 6$ unit cells, 144 phosphorous atoms) is shown for reference. Fixed embedding spaces of 265 average embedding orbitals are used for all nanosheets. (b) Log-log plot of key computation timings vs. system size. ibDET is broken down into bath construction (1-PNO construction and integral transformations) and impurity solver steps. Power law fits are shown as dashed lines. 
    }
    \label{fig:3}
\end{figure}

We first examine the performance of ibDET for modeling nanomaterial systems:
a silicon nanocluster and a series of phosphorene nanosheets.
HF was used as the low-level theory and the $G_0W_0$@HF approach was used as the impurity solver.  The purpose here is to carefully examine the convergence of ibDET with respect to embedding size and to analyze the computational scaling for large systems of varying sizes, for which the full-space limit at the 
$GW$ level is much less expensive than at the EOM-CCSD level.
Results of HOMO and LUMO quasiparticle energies obtained from ibDET compared to the full-space $G_0W_0$@HF are shown in Fig.~\ref{fig:2} and Fig.~\ref{fig:3}.

We first assessed the accuracy of a silicon nanocluster  Si\textsubscript{32}H\textsubscript{44}, which in our computational basis has 1704 basis functions. We ran four ibDET calculations for a series of 1-PNO threshold values from $1.0\times10^{-3}$ to $4.0\times10^{-5}$, with these thresholds giving average embedding space sizes from 103 to 309 orbitals, respectively. Both the HOMO and LUMO quasiparticle energies converge quickly with respect to the embedding space. To achieve IP/EA values from embedding that closely match the full-space results, we devise an extrapolation scheme based on the linear relationship between the calculated IPs/EAs and the inverse square of the number of occupied embedding orbitals $\tilde{N}_\text{occ}$. To conduct a single linear fit of the full system values, we take the average $\tilde{N}_\text{occ}$ over the multiple fragments in each system. As shown in Fig.~\ref{fig:2}c and Fig.~\ref{fig:2}d, our extrapolated results near exactly match the full-space $GW$ results (HOMO and LUMO errors of 0.010 eV and 0.028 eV respectively). The density of states corresponding to the largest embedding space is of similar high quality, almost perfectly matching the full-space reference.

\par 
Next, we tested the accuracy and computational efficiency of our molecular ibDET implementation on increasingly larger 2D phosphorene nanosheets\cite{wilhelmLowScalingGWBenchmark2021}. We considered sizes from $3\times3$ to $6\times6$ unit cells with hydrogen termination, which correspond to 1504 and 5484 basis functions, or 36 and 144 phosphorus atoms, respectively (Fig.~\ref{fig:3}). In all sheets, each impurity problem consists of local IAO+PAO orbitals centered on a phosphorus atom and possibly neighboring hydrogen atoms. We forego our extrapolative error scheme to focus primarily on analyzing the computational complexity of ibDET. Still, even without extrapolation, ibDET predicts HOMO and LUMO quasiparticle energies within 0.2 eV of the full-space results at only 265 average embedding orbitals per impurity (Fig.~\ref{fig:3}a). However, we do observe increasing embedding errors as the sheet size increases, likely due to accumulation of self-energy assembly errors and less localized electron correlation in larger nanosheets. In terms of complexity, we expect each embedding orbital construction step to have roughly cubic scaling with respect to the full system size, coming from computation of the 1-PNO amplitudes ($t_{i\tilde{j}}^{\tilde{a}\tilde{b}}$ and $t_{\tilde{i}\tilde{j}}^{a\tilde{b}}$) and the final rotation of the ERI to the embedding space (Fig.~\ref{fig:1}a). 
The impurity $G_0W_0$@HF calculation thus becomes a small constant cost for each impurity problem, i.e., $\mathcal{O}(1)$, compared to the embedding orbital construction in ibDET. When considering the cost for all $N_\mathrm{atom}$ impurities, the bath constructions have an overall scaling of $\mathcal{O}(N^4)$, while the impurity $G_0W_0$@HF Green's function calculations scale as $\mathcal{O}(N)$. Though this $\mathcal{O}(N^4)$ is not a formal improvement over the canonical full-space $G_0W_0$ scaling, ibDET achieves significant gains in terms of prefactors. 
Both full-space and ibDET calculations exhibit an empirical scaling law of $\mathcal{O}(N^{3.9})$, while ibDET is faster by $2\times$ (parallel shift on log-log scale). More encouragingly, for any MBGF theory more costly than $G_0W_0$ (e.g., CCGF), we expect more substantial formal complexity advantages over full-system calculations. In such cases, the integral transformations carry relatively small cost compared to the impurity solver.

\subsection{Conjugated Molecules with Coupled-Cluster Impurity Solver}
We then benchmarked ibDET with a CCGF impurity solver (at EOM-CCSD level) for two medium-sized conjugated molecules in a cc-pVDZ basis set. We calculated impurity self-energy on a dense real-frequency grid and perform a full Dyson inverse to obtain final MBGF and DOS. From the final ibDET DOS, we fit the HOMO/LUMO peaks to a Lorenzian function and obtain their real-axis positions, which we report here as the HOMO and LUMO quasiparticle energies. 

\begin{figure}[hbt!]
    \centering
    \includegraphics[width=0.8\linewidth]{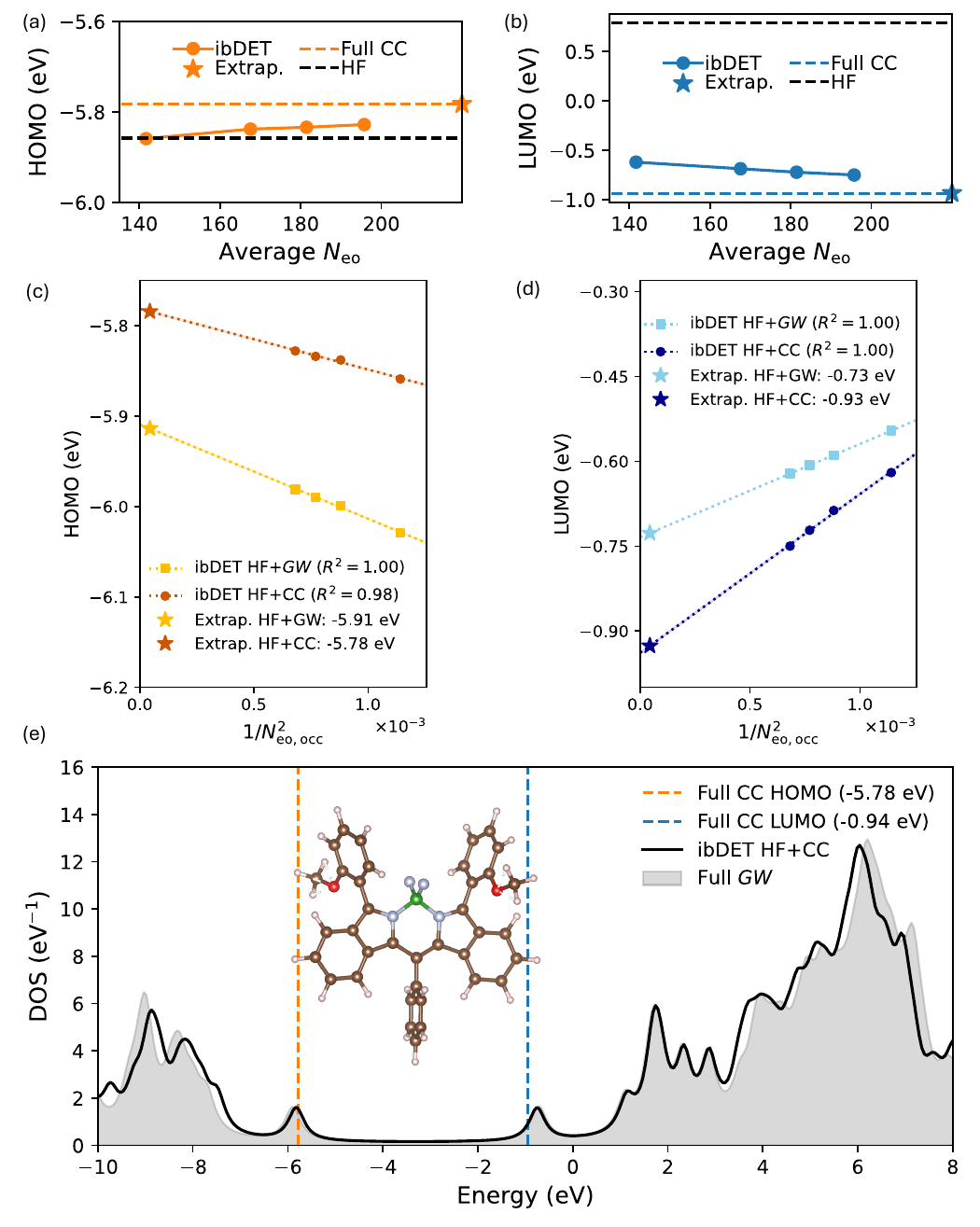}
    \caption{BODIPY molecule HF+CC ibDET benchmark. (a,b) HOMO, LUMO energies vs.~embedding size, with Hartree-Fock and full IP/EA-EOM-CCSD for reference. (c,d) Extrapolation to the full-space limit with respect to embedding size. Extrapolated values shown on (a) and (b) as a star. (e) ibDET-predicted EOM-CCSD density of states overlaid with IP/EA-EOM-CCSD values shown as dashed line. Full-space $G_0W_0$@HF spectrum is also shown for reference. The full-space $GW$ HOMO and LUMO are $-5.89$ eV and $-0.68$ eV respectively.}
    \label{fig:4}
\end{figure}

\par As our first system, we selected a diaryl-substituted BODIPY derivative, as we believe that its diverse atom types and a mixture of in-plane and out-of-plane aromatic substituents make it a suitably challenging system for quantum embedding (Fig.~\ref{fig:4}). Furthermore, BODIPY derivatives have practical applications as dyes and as fluorophores in biological imaging, motivating the development of predictive tools that can be integrated into rational design efforts \cite{luStructuralModificationStrategies2014}. The full-space of this molecule (Fig.~\ref{fig:1}b) comprises 751 orbitals, making IP/EA-EOM-CCSD calculations a challenge for conventional coupled-cluster solvers. We employed PySCF to obtain the HOMO ($-5.782$ eV) and LUMO ($-0.941$ eV) references, which took two days on a single CPU node with 1 TB of memory and 64 cores. Calculation of the corresponding Green's function over a wide spectral range would put an even greater strain on both software and hardware. Comparing our ibDET HF+$GW$ efforts to ibDET HF+CC, we find accurate prediction of the HF+CC self-energy and downstream HOMO and LUMO energies is indeed more challenging than HF+$GW$, we believe due to the more complicated structure of the EOM-CCSD self-energy. With a 1-PNO threshold of $5.0\times10^{-5}$, the average embedding space is 196 orbitals. HOMO and LUMO energy errors at this size are $0.046$ and $0.191$ eV. These results reflect the inherent difficulty of predicting virtual vs. occupied quasiparticle energies in low-scaling correlated excited-state methods. Embedding-size extrapolation effectively reduces these errors to near-zero: $0.002$ eV error for the HOMO and $0.015$ eV for the LUMO. To more clearly understanding the ibDET errors, we also assessed the HF+$GW$ errors, which show a similar improvement of the HOMO and LUMO energies upon extrapolation. However, the benefit is less than HF+CC, and the HF+$GW$ vs. full-space $GW$ errors are larger, with HOMO and LUMO errors of $-0.091$ and $0.057$ eV respectively. We also show the ibDET-predicted EOM-CCSD DOS, overlaid with the $GW$ DOS and full-space EOM-CCSD reference values (Fig.~\ref{fig:4}e). We note the close agreement with ibDET HF+CC DOS with the shape of full-space $GW$ spectrum.

\begin{figure}[hbt!]
    \centering
    \includegraphics[width=0.8\linewidth]{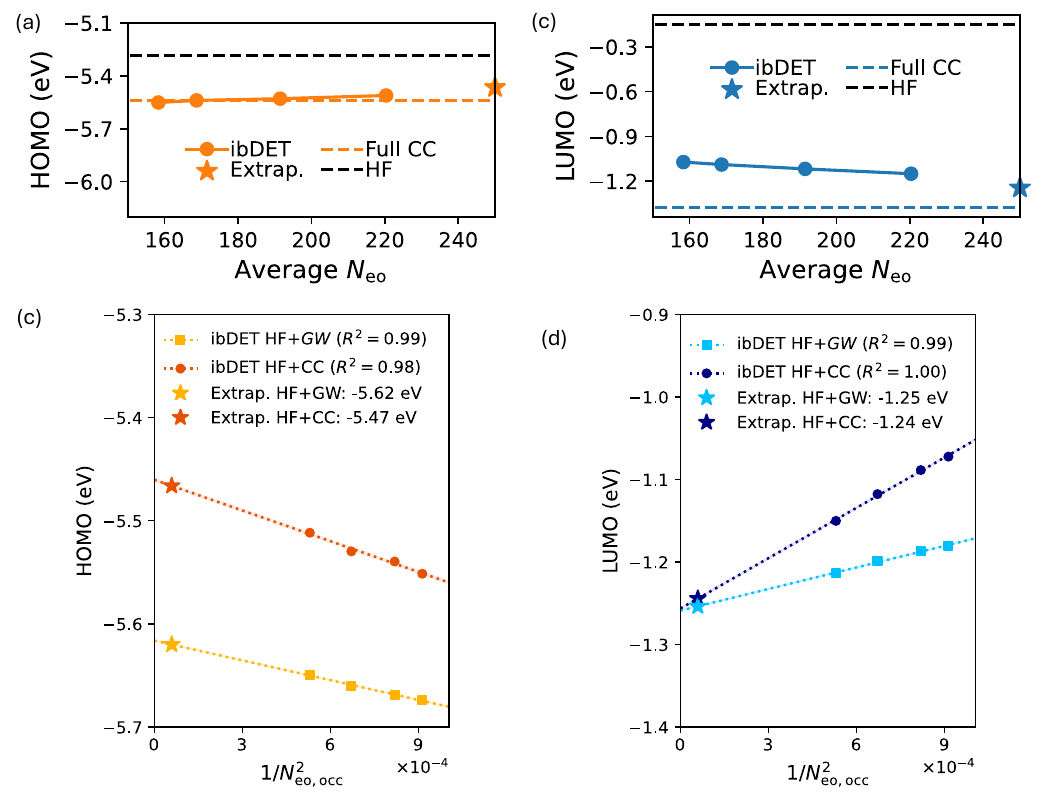}
    \caption{Quaterrylene molecule HF+CC ibDET benchmark. (a,b) HOMO, LUMO energies vs.~embedding size, with Hartree-Fock and full IP/EA-EOM-CCSD for reference. (c,d) Extrapolation to the full-space limit with respect to embedding size. Extrapolated values are shown on (a) and (b) as a star. The full-space $GW$ HOMO and LUMO are $-5.59$ eV and $-1.25$ eV respectively. The full-space IP/EA-EOM-CCSD HOMO and LUMO are $-5.54$ eV and $-1.38$ eV respectively.}
    \label{fig:5}
\end{figure}

\par 
Lastly, we test our method for quaterrylene, a 2D graphene-like sheet (C\textsubscript{40}H\textsubscript{20}) and 660 basis functions (Fig.~\ref{fig:5}). While smaller than realistic carbon nanoribbons, quaterrylene may have qualitatively similar electronic structure and orbitals that make it a useful initial test for ibDET extended to coupled-cluster simulations of large-scale graphene-based systems. For our largest ibDET calculation, we selected a 1-PNO threshold of $5.0\times10^{-5}$ (same as BODIPY), giving an average embedding space of 220 orbitals. While the LUMO energy slowly converges to the full-system reference, the ibDET HOMO energy slightly overshoots the corresponding reference value at a relatively small embedding space of 168 orbitals. Due to this unfavorable convergence behavior, which was absent in our BODIPY benchmark, the HOMO energy error slightly degrades upon embedding-size extrapolation (from 0.028 eV error to 0.074 eV error). The LUMO energy on the other hand benefits from extrapolation, with the embedding error of 0.228 eV at $N_\mathrm{eo} = 220$ reduced to 0.131 eV upon extrapolation. We believe that the diminished performance of ibDET for quaterrylene compared to BODIPY is a result of quaterrylene's more delocalized electron correlation and specific self-energy structure in the IAO space. During ibDET self-energy assembly, off-diagonal errors are more significant and do not favorably cancel, leading to relatively less accurate spectral features. Interestingly, the extrapolative fit is still good ($R^2>0.98$) even when the extrapolated result is of poor quality. Despite the finer details of our ibDET benchmark of quaterrylene, we stress that these $\sim 0.1$ eV errors are still quite small relative to other error sources in correlated excited-state calculations.

\section{Conclusion}
We present a low-cost Green's function embedding approach, interacting-bath dynamical embedding theory (ibDET), to predict accurate spectral properties of molecular systems.
Our method achieves good predictions of IPs and EAs at both the $GW$ and EOM-CCSD levels for a variety of systems with errors around $0.1$ eV or smaller.
Using the new scheme to include natural bath orbitals capturing dominant electron correlation,
quasiparticle energies computed from ibDET converge quickly with respect to the size of the embedding problem,
and results at full-space limit can be obtained with substantially reduced cost.
We demonstrate that our implementation is more practically efficient than full-space calculations, in line with our theoretical complexity analysis.
This work paves the way for scalable correlated Green's function calculations for complex molecular and nanomaterial problems.

\begin{suppinfo}
Calculated IP/EA values for all main-text figures and additional results for BODIPY. All 3D coordinates of molecules.
\end{suppinfo}

\begin{acknowledgement}
This work was primarily supported by the National Science Foundation under Grant No.~CHE-2337991. The quantum embedding software infrastructure development was supported by the National Science Foundation under Grant No.~OAC-2513473. C.V. acknowledges support from the Department of Defense through the National Defense Science \& Engineering Graduate (NDSEG) Fellowship Program. We thank the Yale Center for Research Computing for guidance and use of the research computing infrastructure.
\end{acknowledgement}

\bibliography{ref}

\end{document}